\documentstyle[preprint,aps,epsfig]{revtex}
\def\bea{\begin{eqnarray}}
\def\eea{\end{eqnarray}}
\begin{document}
\draft
\mediumtext
\tightenlines
%\twocolumn[
\title{
Breaking a one-dimensional chain:
fracture in 1 + 1 dimensions
}
\author{
Eugene B. Kolomeisky$^{1}$ and Joseph P. Straley$^{2}$}
\address{
$^{1}$Department of Physics, University of Virginia, Charlottesville, Virginia 22901 
\\
$^{2}$Department of Physics and Astronomy, University of Kentucky,Lexington, Kentucky 
40506-0055 
}
\maketitle
\begin{abstract}
The breaking rate of an atomic chain stretched at zero 
temperature by a constant force can be calculated in a 
quasiclassical approximation by finding the localized solutions 
("bounces") of the equations of classical dynamics in imaginary 
time.  We show that this theory is related to the critical cracks 
of stressed solids, because the world lines of the atoms in the 
chain form a two-dimensional crystal, and the bounce is a crack 
configuration in (unstable) mechanical equilibrium.  Thus the 
tunneling time, Action, and breaking rate in the limit of small 
forces are determined by the classical results of Griffith.  For 
the limit of large forces we give an exact bounce solution that 
describes the quantum fracture and classical crack close to the 
limit of mechanical stability.  This limit can be viewed as a 
critical phenomenon for which we establish a Levanyuk-Ginzburg 
criterion of weakness of fluctuations, and propose a scaling 
argument for the critical regime.  The post-tunneling dynamics is 
understood by the analytic continuation of the bounce solutions 
to real time.

\end{abstract}
\section{Introduction}

A stretched solid breaks instantaneously when the external 
widening stress exceeds a critical value corresponding to the 
limit of mechanical stability of the system.  For smaller 
stresses the fracture is time-delayed and assisted by thermal 
fluctuations.  At sufficiently low temperatures there can be 
quantum effects, and the breaking rate becomes temperature 
independent.  This last situation is the main subject of this 
paper.

We consider a one-dimensional system (for example, a polymer 
chain or nanotube) stretched by a constant external force applied 
at the chain ends at zero temperature, where quantum tunneling is 
the only mechanism responsible for the fracture.  There have been 
several attempts in the past to study this 
problem\cite{1}\cite{2} however its many-body nature was 
not taken into account until the pioneering work of 
Dyakonov\cite{3}, who gave qualitatively correct results in 
the limits of weak and large external forces.  A more precise 
analytical approach to the problem valid for small external 
forces has been suggested recently by Levitov {\it et al}\cite{4}.

We will adopt a general formalism that allows us to look at 
various limits from a single viewpoint.  According to 
Feynman\cite{5} the tunneling rate per unit length is given by 
$w \propto  |\sum \exp(-A/\hbar )|^{2}$ where {\it A} is the imaginary-time Action 
calculated
along particle trajectories connecting the "initial" (stretched)
and "final" (broken) states of the chain, $\hbar $ is Planck's constant,
and the summation is performed over all trajectories.  In the
quasiclassical limit the trajectories contributing most to the
sum satisfy the stationarity condition $\delta A = 0$.  These
trajectories describe how the system traverses the classically-
forbidden region, going from an unbroken chain into a
configuration from which even a classical chain released at rest
would break irreversibly.  It is convenient to combine this
motion with its time reversal to form a localized path called a
"bounce"\cite{6}.  In the quasiclassical approximation the
tunneling rate is given by
\bea
w \cong  B \exp (- A_{bounce}/\hbar )
\label{Eq1}
\eea
where the amplitude {\it B }and action $A_{bounce}$ are calculated for the
bounce (there is no factor of 2 in the exponential from the
squaring because the bounce covers both the past and future).  We
will demonstrate below that the problem of finding bounce
solutions is identical to the classical problem of finding the
equilibrium crack configuration for a two-dimensional solid
stretched by a constant uniaxial stress -- yet another example of
the correspondence between one-dimensional quantum field theory
and two-dimensional statistical mechanics\cite{7}.  In view of
this equivalence the results of \cite{3} and \cite{4} are
already implied by the classical works of Griffith\cite{8}.

The present contribution to this field contains two main results.  
First, in the limit of a large applied force close to the limit 
of mechanical stability we find an exact bounce solution; 
simultaneously this for the first time determines the equilibrium 
crack configuration in the large stress classical limit.  
Establishing the range of applicability of this solution reveals 
that fluctuations inevitably come into play in the immediate 
vicinity of the limit of mechanical stability.

Second, we show how the bounce solutions can be used to provide 
insight into the real time post-tunneling dynamics.

\section{ General formalism}

The imaginary-time Action describing a one-dimensional chain 
stretched by an external force {\it p }applied to the chain ends has 
the form
\bea
A = \int dx dt [{{\rho} \over {2}} (\dot{u}^{2} + c^{2} u^{\prime 2})
- p u^{\prime} ] +
\int dt [V(h) - p h]
\label{Eq2}
\eea
where $\rho $ is the linear particle density, $x$ is the spatial 
coordinate (along the chain length), $t$ is the imaginary time 
coordinate, $u(x,t)$ is the particle displacement field, and dots
and primes stand for the imaginary time (t), and spatial (x)
derivatives.

The first integral is over all positions and times except for a 
small region near $x = 0$, where the break will appear.  This
separates the line into two pieces, so that for all $t$ the field {\it u}
is discontinuous: $u(x = + 0, t) - u(x = - 0, t) = h(t)$; however,
we expect $h(t)$ to be small outside a well-defined time interval.
We have defined $h(t)$ so that the distance between the two parts
of the chain is $a + h$, where {\it a} is the equilibrium $(p = 0)$
interparticle spacing in the chain.  The coupling between the
segments of the chain in the tear region is given by the
potential of cohesive forces {\it V}({\it h}) which describes the interaction
between the two ends of the broken chain for arbitrary separation
-- that is, it goes beyond the harmonic approximation.

The properties of the function {\it V}({\it h}) can be summarized as follows 
(Fig. 1).  For $|h| \ll  a$ it obeys Hooke's law: $V(h) = \rho c^{2}h^{2}/2a$, so
that the two parts of the chain are joined into one elastic
medium.  For large negative {\it h}, the cohesive potential increases
without bound, $V(h \rightarrow  -\infty ) \rightarrow  + \infty $, reflecting the 
impossibility of
indefinite compression.  For large positive {\it h} it approaches a
constant, $V(h \rightarrow  +\infty ) = 2\gamma $, which is the work done in 
infinitely
separating the halves of the chain; $2\gamma $ can also be called the
bond energy and the factor of 2 is introduced to indicate that
there is an energy $\gamma $ associated with each free end of the half-
infinite chain pieces.  In what follows we will also need to know
the properties of the cohesive force function $G(h) = dV/dh$ shown
schematically on Fig. 2: for small {\it h} it grows linearly with $h,
G(h) = \rho c^{2}h/a$, reaching a maximal value $p_{c} \cong  \rho c^{2} \cong  
\gamma /a$ at $h = h_{c}
\cong  a$, while {\it G}({\it h}) decreases to zero as $h \rightarrow  +\infty $.  
The parameter $h_{c}$
can be called the critical bond lengthening, while $p_{c}$ is the
limit of mechanical stability of the system -- indeed only for $p
< p_{c}$ does $U(h) = V(h) - ph ($the total potential energy in the
external field) have a metastable minimum at $h = h_{1} ($Fig.3)
corresponding to the equilibrium stretched chain.

The bounce solution is shown schematically on Fig.4 using the 
imaginary time variable $t$ as a second (space-like) coordinate.  
The particle world lines belonging to the two ends of the 
breaking chain deviate significantly from each other for a time 
interval and then come back together.  The external force {\it p 
}applied at the chain ends is time-independent, therefore it is 
shown schematically by series of arrows of the same length 
pointing in the spatial (x) direction.

The Action (\ref{Eq2}) and the Figures can be interpreted as 
representing a two-dimensional "crystal" of particle world lines 
subject to a uniaxial stress {\it p}, and in this interpretation the 
bounce of Fig. 4 is a critical crack, poised between the small 
perturbations that can spontaneously heal and the large 
disruptions that lead to fracture.  We note in passing that our 
point of view that the crack opening is nonzero everywhere is 
different from that of many classical approaches\cite{9}, 
which restrict the break to a finite region completely surrounded 
by an elastic medium.  Our treatment contains the standard 
approach as a special case.

To see the connection to the classical fracture problem 
explicitly let us seek extremal paths for the Action (\ref{Eq3}) by 
varying $u$ while keeping the boundary values $h(t)$ fixed.  The 
condition $\delta A/\delta u = 0$ reduces to the Laplace equation
\bea
d^{2}u/dt^{2} + c^{2} u^{\prime \prime} = 0
\label{Eq3}
\eea
The solution to (\ref{Eq3}) satisfying the boundary conditions $u(x
= \pm  0, t) = \pm  h(t)/2$ has the form
\bea 
u(x,t) = {p\over \rho c^{2}} x + {{sign x}\over 2} \int^{+\infty }_{-\infty 
}{d\omega \over 2\pi } h(\omega )\exp[i\omega t - (|\omega ||x|/c)]
\nonumber\\
\equiv  {p\over \rho c^{2}} x + {x\over 2\pi c} \int^{+\infty }_{-\infty } {dt^\prime  
h(t^\prime )\over (t^\prime  - t)^{2} + x^{2}/c^{2}}
\label{Eq4}
\eea
where the first term describes the stretched chain, the second is 
due to the break, and the Fourier transform $h(\omega ) =
\int h(t)\exp(-i\omega t) dt$ has been introduced.  Substituting (\ref{Eq4}) in
(\ref{Eq2}), integrating over x, and ignoring overall constants we
find the Action in a form that depends only on the field {\it h}
\bea
A = -{D\over 2}\int^{}_{|t-t^\prime |\ge t_{0}}\dot{h}(t) dt
\ln (|t - t^\prime |/t_{0}) \dot h (t^\prime )dt^\prime  + \int dt[V(h) - ph]
\label{Eq5}
\eea
where $D = \rho c/2\pi $ and $t_{0} \cong  a/c$ is a cutoff introduced to prevent
singularities for $t \rightarrow  t^\prime $.  If imaginary time is viewed as a
space-like variable then the Action (\ref{Eq5}) can be interpreted
as a Hamiltonian describing a "crack" in a two-dimensional
"crystal" of particle world lines.  The crack can be represented
by a distribution of fictitious dislocations\cite{9}, and then
the first term of (\ref{Eq5}) describes the logarithmic interaction
of these dislocations with each other with an interaction
strength {\it D}, a "dislocation density" given by $ -\dot h $, and "Burgers
vectors" pointing along the chain.  Similarly the second integral
in (\ref{Eq5}) can be thought of as describing the cohesive
interaction {\it V}({\it h}) between the sides of the "crack" in the presence
of the external opening field {\it p}.  A free-energy functional
similar to (\ref{Eq5}) has been proposed in \cite{10} and
\cite{11} as a field-theoretical starting point of classical
fracture mechanics.

The configuration $h(t)$ for which (\ref{Eq5}) is extremal is 
determined by the condition $\delta A/\delta h = 0$:
\bea 
D \int^{+\infty }_{-\infty }{{\dot h (t^\prime ) } \over {t^\prime  - t}} 
dt^\prime = G(h) - p
\label{Eq6}
\eea
Here and below the singular integral is taken as a principal 
value.  A model of cracks based on a singular nonlinear integral 
equation of the form (\ref{Eq6}) has been analyzed by Blekherman 
and Indenbom\cite{12}.

The sound velocity c sets the upper limit on the value of $|\dot h|$, 
therefore the physical solutions to (\ref{Eq6}) should satisfy 
$|\dot h (t)/c| \le  1$; at the same time the field $h(t)$ can take on
arbitrarily large values.

A bounce solution to (\ref{Eq6}) will have the following form 
(Figs. 3, 4).  For $t \rightarrow  -\infty $ the function $h(t)$ starts at the
metastable minimum of the total potential $U(h), h = h_{1}$, then
makes an excursion past the unstable maximum, $h = h_{2}$, and comes
back to $h = h_{1}$ as $t \rightarrow  +\infty $.

Substituting (\ref{Eq6}) into (\ref{Eq5}) and subtracting from the 
result the Action of the stretched unbroken chain $(h = h_{1})$, we
find the tunneling Action that goes into the probability
(\ref{Eq1})
\bea
A_{bounce} = \int^{+\infty }_{-\infty }dt[U(h) - U(h_{1}) - {h - h_{1}\over 2} 
{dU(h)\over dh}]
\label{Eq7}
\eea
where we used the result due to Blekherman and 
Indenbom\cite{12} that $\int^{+\infty }_{-\infty }dt[G(h) - p] = 0$ (total zero force 
along the tear) for the solution to (\ref{Eq6}) satisfying $h(\pm \infty ) =
h_{1}$.

For large $|t|$ we can use a harmonic approximation, since the 
function $h(t)$ is close to its asymptotic limit $h_{1}$.  Thus we
introduce $\varphi  = h - h_{1}$ and approximate the right-hand side of
(\ref{Eq6}) by $U^{\prime\prime}(h_{1})\varphi $, valid for $|\varphi | \ll  a$.  
Expanding the integrand
of (\ref{Eq6}) in $t^\prime /t$ and noting that $h(t)$ is even in $t$ we find
\bea
\varphi (t \rightarrow  \pm \infty ) = {D\over t^{2}U^{\prime\prime}(h_{1})} 
\int^{+\infty }_{-\infty }\varphi (t^\prime )dt^\prime 
\label{Eq8}
\eea
However, this is not very interesting, since within the harmonic 
approximation the tunneling Action (\ref{Eq7}) vanishes.

Further progress can be made by looking separately at the cases 
of weak and strong tearing force where the tunneling Action can 
be computed in a controlled fashion.

\section{Small-force limit, p$ \ll$ p${}_{c}$}

In studying the weak force limit, $p \ll  p_{c}$, we can import some 
ideas from classical fracture mechanics{\bf .  }The position of the 
unstable maximum $(h = h_{2})$ of the function {\it U}({\it h}) (Fig.3) shifts to
infinity as $p \rightarrow  0$.  However the amplitude of the bounce solution
must be larger than $h_{2}$, thus implying that over some time the
distance between the edges of the tear is much bigger than
equilibrium interparticle spacing.  Since the bounce tail
(\ref{Eq8}) does not contribute to the tunneling Action (\ref{Eq7})
in quadratic order, we can take {\it h} to be nonzero only within a
time interval $-L < t < L ($Fig.4).  The tunneling time 2L will be
determined later.  The cohesive force, {\it G}({\it h}), is negligible for $-L
\le  t \le  L$ and operative beyond this interval.  The latter feature
implies that there is a strain outside the segment $[-L; L]$ that
takes care of the assumed $h(t) = 0$.  Then (\ref{Eq6}) simplifies to
\bea
D \int^{+L}_{-L} {{\dot h (t^\prime )} \over {t^\prime  - t}} 
dt^\prime = - p
\label{Eq9}
\eea
for $-L \le  t \le $ L.  Neglect of the cohesive forces within the
tunneling interval implies that the edges of the tear are force
free, i.e. $u^\prime (x = \pm 0, t) = 0$; this condition applied to the
representation (\ref{Eq4}) with the limits of the integration set
by $-L$ and $L$ indeed reproduces (\ref{Eq9}).

In the same approximation the total potential $U(h) = V(h) - ph$
can be replaced by $U(h) = 2\gamma  - ph$, and (\ref{Eq7}) simplifies to
\bea
A_{bounce} = \int^{+L}_{-L}dt(2\gamma  - {1\over 2} ph)
\label{Eq10}
\eea
where for $p \ll  p_{c}$ we neglected $U(h_{1}) + ph_{1}/2 = - p^{2}a/\rho c^{2}$ 
compared
to $2\gamma $.  The solution to (\ref{Eq9}) satisfying $h(\pm L) = 0$ is the
elliptical crack of fracture mechanics\cite{13} \cite{14}:
\bea
h(t) = {p\over \pi D} \sqrt{L^{2} - t^{2}} \equiv  {2p\over \rho c} \sqrt{L^{2} - 
t^{2}}
\label{Eq11}
\eea
Substituting this into (\ref{Eq10}) and evaluating the integral we 
find
\bea
A_{bounce}(L) = 4\gamma L - p^{2}L^{2}/4D
\label{Eq12}
\eea
This expression determines the Action parameterized by the time 
$2L$, which is chosen so that $A_{bounce}(L)$ is extremal: then the
Action is stationary with respect to all variations in the path
connecting broken to unbroken configurations.  Eq. (\ref{Eq12}) has
a maximum at
\bea 
2L_{G} = 16\gamma D/p^{2} \equiv  8\gamma \rho c/\pi p^{2}
\label{Eq13}
\eea
which is identified as the tunneling time.  In the context of 
classical fracture mechanics this is called the Griffith 
criterion\cite{8} \cite{9}.  Combining (\ref{Eq1}), 
(\ref{Eq12}), and (\ref{Eq13}) we find that for $p \ll  p_{c}$ the fracture
rate (per unit time and per unit length of the chain) is given by
\bea 
w \cong  (1/cL_{G}^{2}) \exp[- A_{bounce}(L_{G})/\hbar ]
\cong  (p^{4}/\gamma ^{2}D^{2}c) \exp(-16\gamma ^{2}D/\hbar p^{2})  
\nonumber\\
\cong 
(p^{4}/\gamma ^{2}\rho ^{2}c^{3}) \exp(- 8\gamma ^{2}\rho c/\pi \hbar p^{2})
\label{Eq14}
\eea
where the prefactor {\it B }from (\ref{Eq1}) has been estimated by 
arguing that the break can occur anywhere in two-dimensional 
space-imaginary time: the prefactor $1/cL_{G}^{2}$ is the density of
{\it independent} bounces that can fill up two-dimensional space-
imaginary time (to be independent, two bounces must be separated
in time by more than $L_{G}$ and in space by more than $cL_{G})$.

The effective mass involved in the tunneling can be estimated 
from an argument parallel to that of Dyakonov\cite{3}.  The 
representation (\ref{Eq4}) implies that the effect of an 
inhomogeneity of size $L_{G}$ in the t-direction perturbs the picture
of the world lines over a distance of order $cL_{G} ($in the x-
direction) containing about $cL_{G}/a$ atoms.  Each of them has a mass
$m = \rho a$, therefore the effective mass $M$ involved in the tunneling
is found to be
\bea 
M \cong  m (cL_{G}/a) \cong  m (\rho c^{2}\gamma /ap^{2}) \cong  m (p_{c}/p)^{2}
\label{Eq15}
\eea
The tunneling time $2L_{G}$ (\ref{Eq13}), the maximal width of the tear
[see Eq(\ref{Eq11})], $h(0) = (p/\pi D)L_{G} = 8\gamma /\pi p$, and the tunneling mass
(\ref{Eq15}) all diverge as $p \rightarrow  0$ implying that fracture through
tunneling has many-body nature.

Here as in fracture mechanics this approach breaks down in the 
vicinity of $t = \pm  L$ where the derivative $\dot h (t)$ diverges.  There is
also an associated stress singularity (which we will not discuss
here) near $t = \pm L$.  These defects can be remedied by taking
proper account of the cohesive forces\cite{9} \cite{14}
with the general conclusion that all the scaling dependencies
predicted in the framework of the Griffith approximation are
correct and that the numerical factors entering Eqs.(\ref{Eq12})-
(\ref{Eq14}) are accurate in the limit $p \ll  p_{c}$; these comments apply
equally well to the quantum fracture problem.

For example, the range of applicability of Eq(\ref{Eq11}) can be 
found by requiring that both $|\dot h| \le  c$ and $h \gg  a$ with the
conclusion that  {\it independently of the value of the tearing force
p}, (\ref{Eq11}) can be trusted outside a small region -- a dozen
microscopic time spans of order $a/c$ in the vicinity of $t = \pm  L_{G}$.

For $p \ll  p_{c}$ the tail of the true solution (\ref{Eq8}) becomes
\bea
\varphi (t \rightarrow  \pm \infty ) = {apL_{G}^{2}\over \rho c^{2}t^{2}} \cong  a 
(p/p_{c}) (L_{G}/t)^{2}
\label{Eq16}
\eea
which is generally small since $p \ll  p_{c}$.

The analysis given above shows that for weak breaking forces $(p \ll 
p_{c})$ the true bounce solution is approximated quite well by
Eq(\ref{Eq11}) for $|t| < L$, and by Eq.(\ref{Eq16}) for $|t| > L$, outside
of the vicinity of $|t| = L = L_{G}$.  The determination of the size
of the transient region (expected to be a dozen microscopic time
scales $a/c)$ and the precise functional form $h(t)$ inside it
require a detailed knowledge of the cohesive potential {\it V}({\it h}); it
should not affect the main conclusions (\ref{Eq13})-(\ref{Eq15}).

Apart from numerical and preexponential factors, the results 
(\ref{Eq13}) and (\ref{Eq14}) were first found by Dyakonov\cite{3} 
using heuristic arguments based on the tunneling properties of a 
particle of effective mass (\ref{Eq15}) passing through a 
triangular potential barrier.  A conformal mapping technique 
similar to what is employed in fracture mechanics\cite{14} was 
used by Levitov {\it et al}\cite{4}; this reproduced Eq(\ref{Eq11}) and 
(\ref{Eq12}), the numerical factors entering in Eq.(\ref{Eq13}) and 
the exponential of Eq.(\ref{Eq14}).  However, the link to classical 
fracture mechanics has not been previously noticed.

Experimental verification of the weak tension results will be 
hindered by the exponentially large lifetime of the chain [see 
Eq.(\ref{Eq14})].  Therefore the case of large tearing force ({\it p 
}close to $p_{c})$, where the lifetime is shorter, seems more important
from the practical standpoint.

\section{ Vicinity of the limit of mechanical stability, $\Delta p \ll  p_{c}$}

For $\Delta p = p_{c} - p \ll  p_{c}$ the tunneling still has a collective nature, 
because the limit of mechanical stability, $p = p_{c}$, is formally
similar to the spinodal of magnetic systems\cite{10}
\cite{11}.  In this limit the tunneling involves the vicinity
of $h = h_{c} ($Fig.2) and the bounce amplitude is very small.
Regardless of the underlying interactions the cohesive force can
be approximated here as $G(h) = p_{c} - b (h - h_{c})^{2}$ where $b = - {1\over 2}
{d^{3}V(h = h_{c})/ {dh^{3}}}$.  The zeros of $G(h) - p = \Delta p - b (h - 
h_{c})^{2}$ determine
the positions of the metastable minimum $(h_{1})$ and unstable maximum
$(h_{2})$ of the total potential energy function {\it U}({\it h}) (see Fig. 3):
\bea
h_{1,2} = h_{c} \mp  \sqrt{\Delta p/b}
\label{Eq17}
\eea
Expressing $G(h) - p$ in terms of $\varphi  = h - h_{1}$ we find
\bea
G(h) - p = 2\sqrt{\Delta pb} \varphi  - b \varphi ^{2}
\label{Eq18}
\eea
Similarly the potential energy function $U(h) = V(h) - ph$ can be
written as
\bea 
U(h) = U(h_{1}) + \sqrt{\Delta pb} \varphi ^{2} - {b\over 3} \varphi ^{3}
\label{Eq19}
\eea
and Eqs. (\ref{Eq6}) and (\ref{Eq18}) give the equation for the 
bounce:
\bea 
D \int^{+\infty }_{-\infty }{\dot{\varphi }(t^\prime )dt^\prime \over t^\prime  - t} 
= 2\sqrt{\Delta pb} \varphi  - b \varphi ^{2}
\label{Eq20}
\eea
Close to the limit of mechanical stability the properties of 
strongly stretched unbroken chain substantially deviate from 
those of the equilibrium one:  the linear mass density $\rho $ is 
significantly reduced, and the lattice looses its stiffness at $p
= p_{c}$.  The latter means that both the sound velocity c, and the
parameter $D = \rho c/2\pi $ entering (\ref{Eq20}) {\it vanish} as $\Delta p \rightarrow  
0$.  The
functional dependencies can be recovered by comparing the
harmonic terms of (\ref{Eq19}) and (\ref{Eq2}), $\sqrt{\Delta pb} \cong  \rho c^{2}/a$, which
gives us
\bea 
D \cong  (m^{2}b\Delta p)^{1/4}, c \cong  a (b\Delta p/m^{2})^{1/4}
\label{Eq21}
\eea
The solution to Eq.(\ref{Eq20}) is
\bea
\varphi (t) = \Phi  {\xi ^{2}\over t^{2} + \xi ^{2}}   ,   
\label{Eq22a}
\eea
where
\bea
\Phi = 4 \sqrt{\Delta p/b}
\label{Eq22b}
\eea
\bea
\xi  = {D\pi \over 2\sqrt{\Delta pb}} \cong  (m^{2}/b\Delta p)^{1/4}
\label{Eq22c}
\eea
Combining (\ref{Eq19}) and (\ref{Eq22a})-(\ref{Eq22c}) we see that the general "sum 
rule" (\ref{Eq8}) is indeed satisfied.  The value of the bounce 
amplitude $\varphi (t = 0) = \Phi $ is small close to $p = p_{c}$; at the same time
$h(0) = \Phi  + h_{1} = h_{c} + 3 \sqrt{\Delta p/b} = h_{2} + 2 \sqrt{\Delta p/b} > 
h_{2}$, i.e. as
expected for the bounce solution, the function $h(t)$ goes past $h_{2}$
for some time.

The divergent time scale $\xi $ (\ref{Eq22c}) can be interpreted as a 
tunneling time, and the effective tunneling mass $M$ can be 
estimated by replacing $L_{G}$ by $\xi $ in (\ref{Eq15}): $M \cong  m (c\xi /a) \cong $ 
m.
We see that as $\Delta p \rightarrow  0$ the effective mass involved in tunneling is
comparable to the single particle mass in agreement with
Dyakonov\cite{3}.

Substituting the potential energy function (\ref{Eq19}) into 
(\ref{Eq7}) with the bounce solution given by (\ref{Eq22a})-(\ref{Eq22c}), and 
computing the integral we find for the tunneling Action
\bea 
A_{bounce} = 2\pi ^{2}D\Delta p/b \cong  (m^{2}/b^{3})^{1/4} (\Delta p)^{5/4}
\label{Eq23}
\eea
which in view of (\ref{Eq1}) implies that for $\Delta p \ll  p_{c}$ the tunneling 
rate is given by
\bea 
w \cong  (1/c\xi ^{2})\exp(-2\pi ^{2}D\Delta p/b\hbar ) \cong 
(b\Delta p/m^{2}a^{4})^{1/4}\exp[ - const (m^{2}/b^{3})^{1/4}(\Delta p)^{5/4}/\hbar ]
\label{Eq24}
\eea
where the prefactor is taken from (\ref{Eq14}), $L_{G}$ is replaced by $\xi$ 
[Eq(\ref{Eq22c})], and the dimensionless constant under the
exponential is of order unity.

Our results (\ref{Eq23}) and (\ref{Eq24}) in the limit $\Delta p \ll  p_{c}$
reproduce those of Dyakonov\cite{3} even though he could not
find the exact bounce solution (\ref{Eq22a})-(\ref{Eq22c}).  Even in the absence
of the exact solution it can be seen that Eq(\ref{Eq20}) has a
solution of the form $\varphi (t) = 4 \sqrt{\Delta p/b} f(t/\xi )$ where $\xi $ is 
given by
(\ref{Eq22c}) and f(z) is a scaling function obeying an integral
equation that has no free parameters at all.  This identifies $\xi  \propto 
(\Delta p)^{-1/4}$ with the tunneling time, and then the tunneling Action
is estimated as proportional to $(\Delta p)^{3/2}(\Delta p)^{-1/4} = (\Delta 
p)^{5/4}$ in
agreement with the calculation that led to (\ref{Eq23}).

In view of the equivalence between the bounce solutions and 
equilibrium cracks of the corresponding two-dimensional classical 
problem, Eqs. (\ref{Eq22a})-(\ref{Eq22c}) also give the equilibrium crack 
configuration in the limit of large stresses close to the limit 
of mechanical stability $p_{c}$.  In this correspondence the imaginary
time t becomes a spatial coordinate along the crack, {\it D} is an
elastic constant\cite{9} [still vanishing according to
(\ref{Eq21})], $ \xi $ is the effective length of the crack, and the
parameter m will have a meaning of a typical binding energy
divided by interparticle spacing.  The rate of the crack
nucleation will be given by (\ref{Eq24}) with the physical
temperature $T$ substituting for the Planck's constant $\hbar $.

The classical problem of finding the equilibrium crack 
configuration in the large stress limit close to the stability 
threshold was previously considered in \cite{11}.  The 
conclusions of that work differ from ours for two reasons.  
First, it was assumed that the long-range nature of the 
interaction between the different parts of the crack profile 
through the bulk of crystal can be replaced by a finite-range 
interaction; as a result the equation for the profile is 
differential rather than integral [see (\ref{Eq20})].  However, 
this is certainly not true for materials obeying conventional 
elasticity theory.  Second, it was assumed that for small $\varphi $,
$G(h)
- p = \Delta p - b\varphi ^{2}$, instead of our Eq(\ref{Eq18}).  But this cannot be
correct, because for $\varphi  = 0 ($metastable equilibrium) the cohesive
force {\it G}({\it h}) balances the external force {\it p} and therefore $G(h) - p$
must vanish.

\section{ Levanyuk-Ginzburg criterion and critical region}

According to (\ref{Eq23}) the tunneling Action vanishes upon 
approach to the limit of mechanical stability $p = p_{c}$.  However,
the results of the previous Sections were based on a
quasiclassical picture which is only valid for $A_{bounce}/\hbar  \gg  1$,
i.e. when the exponential factor of (\ref{Eq1}) dominates the
breaking rate.    This condition imposes a range of validity for
our results
\bea 
\Delta p/p_{c} \gg  (b^{3}\hbar ^{4}/m^{2}p_{c}^{5})^{1/5}
\label{Eq25}
\eea
that eliminates from consideration the immediate vicinity of the 
threshold of mechanical stability $p = p_{c}$.  This condition is
analogous to the Levanyuk-Ginzburg criterion\cite{15} of
weakness for thermal fluctuations in the theory of critical
phenomena.

The results (\ref{Eq21})-(\ref{Eq24}) were also obtained under the 
assumption that $\Delta p/p_{c} \ll  1 [$only then is the expansion in powers
of $\varphi $ used in (\ref{Eq18}) - (\ref{Eq20}) accurate].  This will be
consistent with (\ref{Eq25}) only if
\bea 
{\bf G}{\bf i} \equiv  (b^{3}\hbar ^{4}/m^{2}p_{c}^{5})^{1/5} \ll  1
\label{Eq26}
\eea
The dimensionless quantity {\bf Gi }is a property of a given substance 
and is analogous to the Ginzburg parameter of the theory of 
critical phenomena.  In the present context the requirement ${\bf G}{\bf i} \ll 
1$ can be viewed as the condition of weakness of quantum
fluctuations.

The inequalities (\ref{Eq25}) and (\ref{Eq26}) set the range of 
applicability of the theory of Section IV.  The immediate 
vicinity of $p = p_{c}$ corresponding to the reversed sign in the
inequality (\ref{Eq25}) can be called the fluctuational region;
quantum fluctuations play here a dominant role and the
quasiclassical approximation is insufficient.  In the case of the
reversed sign in the inequality (\ref{Eq26}), ${\bf G}{\bf i} \gg  1$, the theory of
Section IV has no range of applicability at all.

For the case that one of the conditions (\ref{Eq25}), (\ref{Eq26}) is 
broken, our analytic approach fails, but some understanding can 
be gained with the aid of a scaling theory constructed in analogy 
with the theory of critical phenomena\cite{16}.  Assume that 
in the critical regime there is a single independent divergent 
time scale $\xi $, the tunneling time, and all other physical
quantities can be expressed in terms of $\xi $.  According to
(\ref{Eq19}), the barrier for tunneling can be estimated as
$(\Delta p)^{3/2}/b^{1/2}$, and therefore the tunneling Action behaves as
$\xi (\Delta p)^{3/2}/b^{1/2}$.  In the critical region the tunneling Action
should be of the order of Planck's constant $\hbar $, so that
\bea 
\xi  \cong  \hbar b^{1/2}/(\Delta p)^{3/2}
\label{Eq27}
\eea
We also expect that the tunneling mass is still of order m which 
implies $c \cong  a/\xi $ for the critical behavior of the sound velocity.
These conjectures are made plausible by two mild self-consistency
checks: (i) the divergence in (\ref{Eq27}) is stronger than
(\ref{Eq22c})) as expected for the critical regime; (ii) the
dependencies (\ref{Eq27}) and (\ref{Eq22c}) match each other on the
border of the critical regime defined by (\ref{Eq25}).

The conditions (\ref{Eq25}) and (\ref{Eq26}) have the classical 
analogs that the results of Section IV are applicable to the 
classical fracture of solids whenever the following inequalities 
are satisfied
\bea 
\Delta p/p_{c} \gg  (b^{3}T^{4}/m^{2}p_{c}^{5})^{1/5}
\label{Eq28}
\eea
\bea
{\bf G}{\bf i} \equiv  (b^{3}T^{4}/m^{2}p_{c}^{5})^{1/5} \ll  1
\label{Eq29}
\eea
Now the constraints (\ref{Eq28}) and (\ref{Eq29}) describe the 
conditions for thermal fluctuations to be sufficiently weak so 
that the critical crack described by (\ref{Eq22a})-(\ref{Eq22c}) is relevant for 
the case of large applied stresses.

Inside the critical region, $\Delta p/p_{c} \ll  (b^{3}T^{4}/m^{2}p_{c}^{5})^{1/5}$, 
thermal 
fluctuations play a dominant role and the mean-field-like 
treatment of Section IV is invalid.  If the Ginzburg parameter 
(\ref{Eq29}) is not small compared to unity, the theory of Section 
IV does not apply.

In the critical regime the analog of (\ref{Eq27}),

\bea
\xi  \cong  Tb^{1/2}/(\Delta p)^{3/2} , 
\label{Eq30}
\eea
gives the dependence of the critical crack length $\xi $ on the 
distance to the instability threshold $\Delta p$.

\section{Real time dynamics}

Coleman\cite{6} notes that the analytic continuation of the 
bounce to real time describes the evolution of the system {\it after 
}the tunneling takes place.  Introducing the physical time $\tau  =$ -it 
into (\ref{Eq11}) we find that for a small tearing force, $p \ll  p_{c}$,
the separation between the pieces of the chain evolves with time
as
\bea 
h(\tau ) = {2p\over \rho c} \sqrt{L_{G} + \tau ^{2}}
\label{Eq31}
\eea
where L$_{G}$ is the true tunneling time (\ref{Eq13}).  Equation 
(\ref{Eq31}) meets our physical expectations: after penetrating 
through the barrier at $t = i\tau  = 0$, the end of the chain starts
moving classically from rest from the escape position $h(0) =
2pL_{G}/\rho c = 8\gamma /\pi p$ which is the turning point of the classical
motion.  The result (\ref{Eq31}) is also precise analytically as in
getting (\ref{Eq11}) the potential energy function {\it U}({\it h}) was
approximated by $2\gamma  - ph$ which is even more accurate in the
classically allowed region than under the barrier.

For small times $(\tau  \ll  L), Eq$.(\ref{Eq31}) predicts a motion with a 
constant acceleration $2p/\rho cL_{G}$ that can be understood as follows.
Immediately after the break each point of the chain excepting a
finite segment near the edge will be in mechanical equilibrium.
Therefore the whole unbalanced force {\it p} is applied to the edge
segment.  The mass of this region is the same as the tunneling
mass (\ref{Eq15}), $M \cong  \rho cL_{G}$, and its acceleration due to the
external force {\it p} is of order $p/M \cong  p/\rho cL_{G}$ in agreement with
(\ref{Eq31}).  This motion breaks the force balance of the region
next to the moving edge segment causing the propagation of
acceleration away from the tear.

To understand the dynamics inside the chain we use the 
expression\cite{14} \cite{4} for the displacement field 
$u(x,t)$ corresponding to the solution (\ref{Eq11}):
\bea 
u(x,t) = {p\over \rho c^{2}} Re[(x + ict)^{2} + (cL_{G})^{2}]^{1/2}
\label{Eq32}
\eea
Analytically continuing this to the real time $\tau = -it$, we find
\bea
u(x,\tau ) = {p\over 2\rho c^{2}} sign x \left ( [(x - c\tau )^{2} + 
(cL_{G})^{2}]^{1/2}
+ [(x + c\tau )^{2} + (cL_{G})^{2}]^{1/2}\right )
\label{Eq33}
\eea
This expression is a solution to the wave equation $\partial ^{2}u/\partial \tau ^{2} 
-
c^{2}\partial ^{2}u/\partial x^{2} = 0$ in the two parts of the broken chain 
satisfying the 
conditions $u(\pm 0,\tau ) = \pm h(\tau )/2$ and $u^\prime (\pm 0,\tau ) = 0$.  It 
describes
sound waves propagating away from the break and leaving behind
virtually unstrained material moving at constant speed.  Not much
is happening for $\tau  < L_{G}$; however, for $\tau  \gg  L_{G}$ the following
picture emerges (by symmetry we only need to consider the $x > 0$
segment).  For $x \ll  c\tau $ the strain $u^\prime $ is small and $u(x,\tau ) \cong  
p\tau /\rho c$,
whereas for $x \gg  c\tau , \partial u/\partial \tau $ is small and $u(x,\tau ) \cong  
px/\rho c^{2}: a$ moving
unstrained region encounters a stretched immobile region at $x =
c\tau $.  In a transition region having the width of order $cL_{G}$ the
matter is beginning to move to become part of the moving region.
The unstrained region set into motion has mass that grows
linearly with time $[M(\tau ) = mc\tau /a = \rho c\tau ]$ and moves at the constant
speed $v = p/\rho c$; the momentum $M(\tau )v = p\tau $ is growing linearly in
time in response to the constant applied force {\it p}.
Figure 5 shows how this soliton-like wave traveling away from the break is
presented in terms of particle trajectories. The break occurred at 
$x = 0$, $\tau = 0$, and the Figure was constructed by adding 
$u(x,\tau)$ [given by Eq.(\ref{Eq33})] to $x$ for each $\tau$.

In the large force limit, $\Delta p \ll  p_{c}$, we analytically continue the 
bounce solution (\ref{Eq21}) to find that the distance between the 
tearing edges of the chain evolves with time as
\bea 
h(\tau ) = h_{1} + \Phi  {\xi ^{2}\over \xi ^{2} -\tau ^{2}}
\label{Eq34}
\eea
Similar to the case of small tearing force, $p \ll  p_{c}$, the initial 
stage of the motion, $\tau  \ll  \xi ,$, described by Eq(\ref{Eq34}) can be 
interpreted as a constant acceleration of the tunneling mass m by 
the unbalanced force $\Delta p$.  However the range of applicability of
(\ref{Eq34}) is rather narrow -- Eqs(\ref{Eq22a})-(\ref{Eq22c}) [and thus (\ref{Eq34})]
were obtained assuming the expansions (\ref{Eq18})) and (\ref{Eq19})
to be valid.  At the same time the distance between the tearing
edges grows with time according to (\ref{Eq34}), thus implying that
as time progresses, $h(\tau ) - h_{1}$ becomes comparable to the
equilibrium interparticle spacing {\it a} and the expansions (\ref{Eq18})
and (\ref{Eq19}) cease to be accurate.  Therefore (\ref{Eq34}) is
valid only for $\tau  \ll  \xi $.  In the same approximation the motion in
the chain can be obtained by substituting (\ref{Eq22a}) in
(\ref{Eq4}), evaluating the integral, and doing analytic
continuation to the real time $\tau  = -it$:
\bea 
u(x,\tau) = {p_{c}\over \rho c^{2}} x + {sign x \over 2} \left ( h_{1} + \Phi \xi  
{{|x|/c + \xi } \over {[|x|/c + \xi ]^{2} - \tau^{2}}} \right )
\label{Eq35}
\eea
where the sound velocity c is given by (\ref{Eq21}).  Beyond the 
range $\tau  \ll  \xi $, the dynamics is expected to be described by a theory
qualitatively similar to the one given above for the case $p \ll  p_{c}$
with $p_{c}$ and $\xi $ substituting for {\it p} and $L_{G}$.

\acknowledgments

Our interest to this subject was initiated by very fruitful 
arguments with A.Buchel and J.P.Sethna.  We also thank K.Burton, 
C.L.Henley, H.Hui, S.L.Phoenix, and A.V.Shytov for critical 
remarks.

\begin{figure}
\centerline {\epsfxsize=3.in\epsfbox{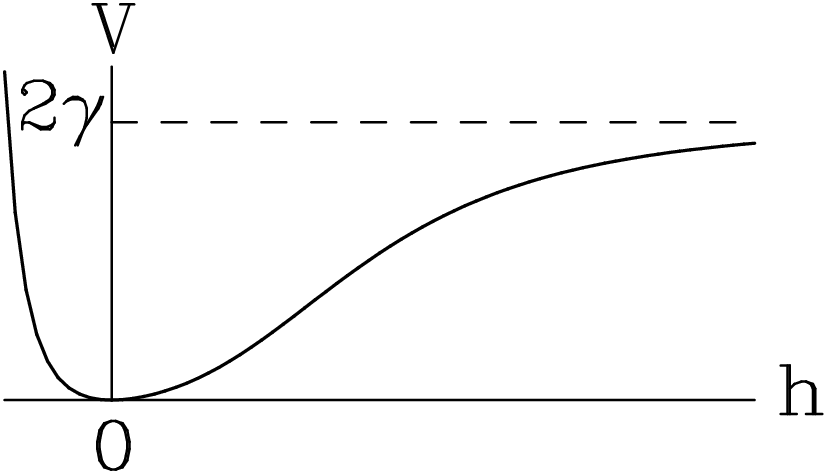}}
\caption{
The cohesive potential {\it V}({\it h}) acting between the edges of the chain 
segments.  For $h \rightarrow  \infty $ it saturates at the value $2\gamma $ 
corresponding
to the bond energy.
}
\end{figure}
\begin{figure}
\centerline {\epsfxsize=3.in\epsfbox{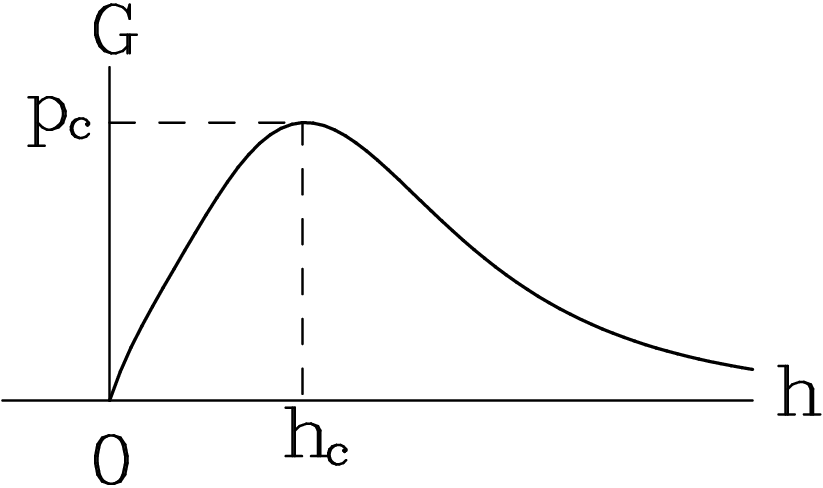}}
\vspace{3mm}
\caption{
The cohesive force $G(h) = dV/dh$. It reaches a maximal value $p_{c}$ at
the critical bond lengthening $h_{c}$ which corresponds to the limit
of the mechanical stability of the system.
}
\end{figure}
\begin{figure}
\centerline {\epsfxsize=2.in\epsfbox{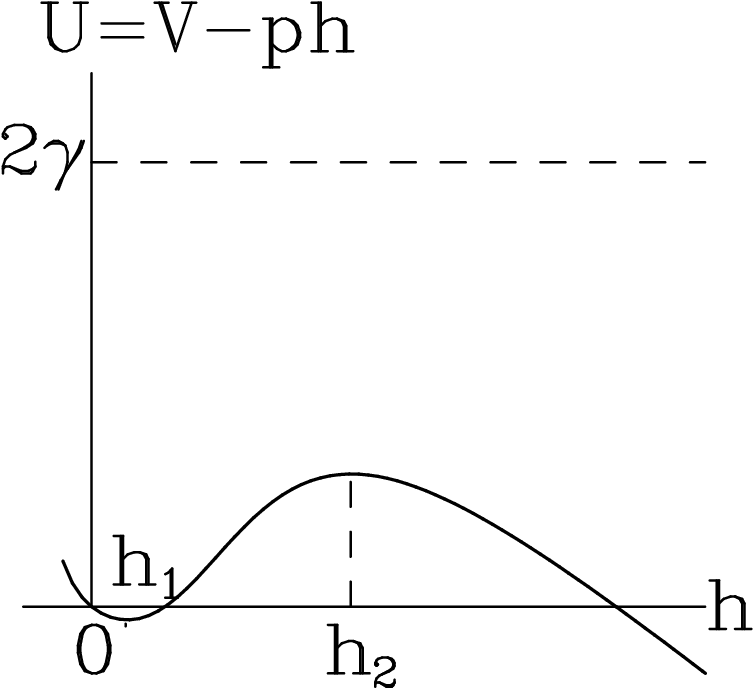}}
\vspace{3mm}
\caption{
The total potential energy $U(h) = V(h) - ph$.  In the presence of
an external tearing force $p < p_{c}$ the total potential energy has a
metastable minimum at $h_{1}$ describing the untorn stretched chain
and an unstable maximum at $h_{2}$, which is the position of the peak
of the tunneling barrier.
}
\end{figure}
\begin{figure}
\centerline {\epsfxsize=2.in\epsfbox{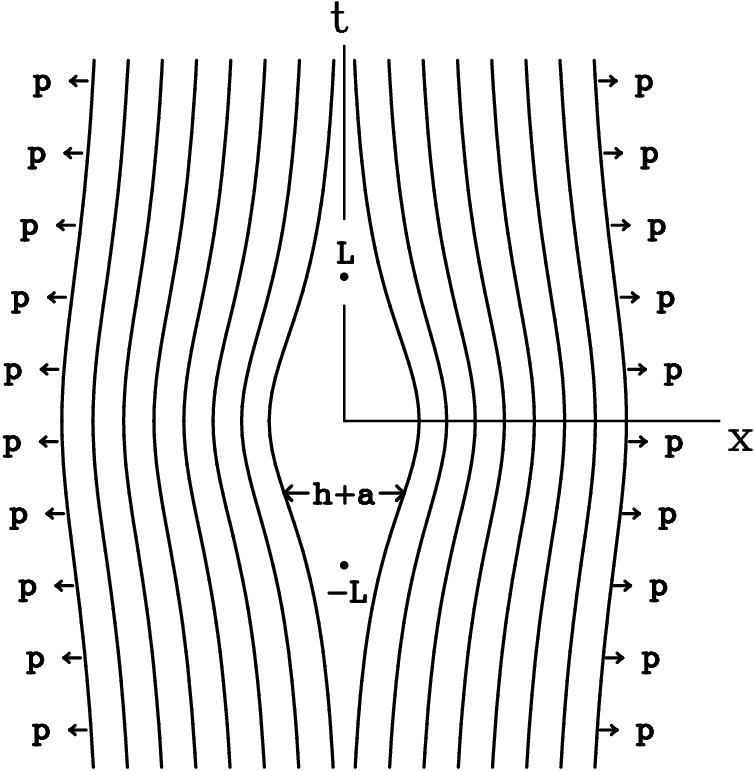}}
\vspace{3mm}
\caption{
The two-dimensional "crystal" of particle world lines torn by a 
constant external force {\it p }traversing the imaginary time (t) 
direction.  The tear takes place at $x = 0$, the distance between
the torn edges is $h + a$, the tunneling time is 2L, and the system
penetrates through the barrier at $t = 0$.
}
\end{figure}
\begin{figure}
\centerline {\epsfxsize=3.in\epsfbox{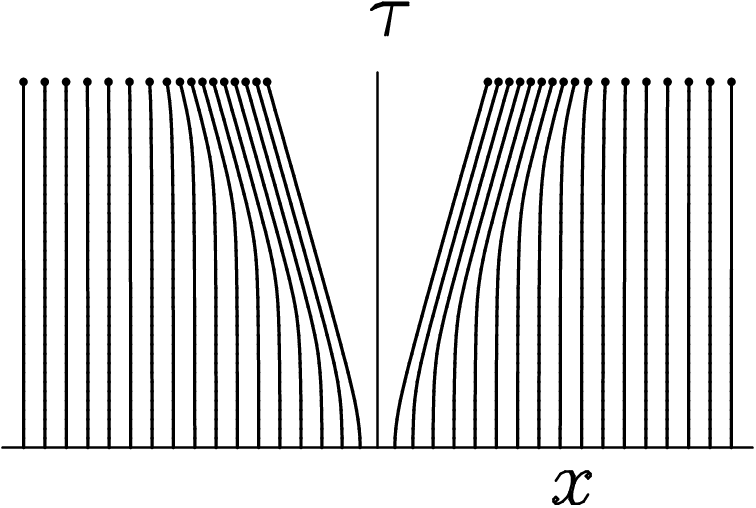}}
\vspace{3mm}
\caption{
The collection of particle trajectories constituting a breaking
one-dimensional solid torn by a constant external force p.
}
\end{figure}
\end{document}